# SYNTHESIS AND CHARACTERIZATION OF BORON NITRIDE POWDERS PRODUCED UNDER CONCENTRATED LIGHT

Lina Sartinska[1], Yevgen Voynich[1], Tarik Eren[2], Esra Altay[2], Oleksandr Koval[1], Izabella Timofeeva[1], Anatoliy Kasumov[1], Gennadiy Frolov[1], Cesarius Jastrebski[3], Vitaly Tinkov [4]

[1] Frantsevich Institute for Problems of Materials Science of NASU,
3 Krzhyzhanovsky Str., Kiev-142, 03680 Ukraine;
[2] Chemistry Department, Faculty of Art & Sciences, Yildiz Technical University, Davutpasa Campus, 34220 Esenler, Istanbul, Turkey;
[3] CEPHOMA Centre at Faculty of Physics Warsaw University of Technology, 75 Koszykowa Str., 00-662 Warszawa, Poland;
[4] Kurdyumov Institute for Metal Physics of NASU,
Vernadsky Blvd. 36, Kiev-142, 03680, Ukraine.

## ABSTRACT

Synthesis and research of the properties of boron nitride powders and BN powders with additives $NiSO_4$ produced under effect of concentrated light in a flow of nitrogen in a xenon high-flux optical furnace are presented.
A scanning and transmission electron microscopes demonstrated structures of new morphologies for the powders, which were formed. X-ray Diffraction study, Raman scattering and electron diffraction study have confirmed a complicated structure and phase composition of the powders with a prevalence of the amorphous phases.
It was demonstrated an effect of experimental conditions, surface modification and additives on phase composition, morphology and structure formation. The "gaseous model" based on an evolution of the bubble has been confirmed new nanostructure formation. Burst of these bubbles may result in graphene-like structures formation.

## INTRODUCTION

Boron nitride (BN) is a synthetic chemical compound containing boron (B) and nitrogen (N) atoms in a one-to-one ratio. Arrangements of nitrogen and boron atoms and their specific bonding behavior result in BN that exists in many different structures. The well-known four polymorphs of BN are cubic (c-BN), hexagonal (h-BN), rhombohedral (r-BN) and wurtzite (w-BN). The variety of interesting properties of boron-nitrogen materials are closely related to their crystal structures.

BN crystal structures have their own analogues of all the carbon forms. Similarity in electronic structure to carbon shares the same number of electrons between adjacent atoms, however bonding character and structural defects are completely different. The two most-used forms of boron nitride are its equivalents of graphite and diamond. The hexagonal (graphite) structure of BN allows the molecules to be arranged in layers[1]. The in-plane atoms are linked through covalent bonds, while the out-of-plane layers are bonded by weak interactions (van der Waals forces) between B and N atoms, alternatively, providing anisotropic properties[2]. h-BN is a wide band gap semiconductor with potential applications in optoelectronic devices that stable at room temperature and ambient pressure.

Another parallel with carbon is the ability to grow boron nitride nanotubes (BNNTs)[3]. BNNTs are very strong and light. Unlike the carbon nanotubes, though, the BNNTs are insulators, less chemically reactive and less susceptible to breakdown in high temperatures[3].

Application of BN nanotubes for the making of optical modulators for ultra-short laser pulses is one of the promising direction of a science. Boron 10 isotope incorporated in BNNTs results in the lighter radiation absorption, meaning that such nanotubes could act as 'radiation shields'[4].

There is also a boron nitride equivalent of the new carbon form, graphene[5]. Sometimes called 'white graphene' these nanosheets of boron nitride are already proving equally versatile. Like graphene, the two dimensional sheets of BN have wide applications in electronics, where the insulating capability make them a natural partner for graphene's conductivity.

Recently, many studies have reported the preparation of nanostructures of the boron nitride with many special morphologies, such as fullerenes[6,3], nanotubes[7,3,8,9], nanocapsules[10], nanocages[11], porous structures[12], hollow spheres[13], nanofibers[14,15], graphene-like structures[16]. All these special morphologies strongly depends on synthesis condition and catalysts[17]. BN nanostructures including nanotubes, nanofibers and nanosheets having a large surface area are very useful in storing hydrogen and other gases[18].

Synthesis forms a vital aspect of the science of nanomaterials. In this context, chemical methods have proved to be more effective and versatile than physical methods for nanomaterials production[19]. However, the physical methods offers a number of significant advantages that have to be exploited too. Among physical methods of synthesis, it is necessary to emphasize heating under concentrated light in a solar simulator. The solar simulator is a unique facility that allows much more flexibility and control over experiments. The simulator is ideal for attaining reproducible results and examining of various processes. The main features of a concentrated light heating are purity, practically absence of inertia of the heating, the relatively high operating temperatures (up to 3000 ºC), high-temperature gradients, and the ability to handle in the air, in vacuum and in the protective, oxidizing or reactive atmospheres. It is local, one-sided heating with radial symmetry of the heating zone[20].

Our research has demonstrated that high-temperature gradients, initial powders, catalysts and nitrogen purification are important to determine the synthesis of boron nitride of special morphologies[21,22]. Therefore, the main idea of this investigation to carry out careful and logical consideration of how evidence of different experimental factors supports or does not support an morphological design, of how different experimental factors are related to one another, and of what sorts of things we can expect to observe if a particular factors is the most effective.

## EXPERIMENTAL

Compacted plate fine-grained powder of boron nitride (Chempur, CH070802) has been used as an initial. The origin powder is h-BN textured on 002 with impurity of $B_2O_3$. The mean size of h-BN plates is about ~ 0,3 μm, their thickness ~ 0,01 μm. In order to increase the chemical stability of h-BN the initial powders were annealed at 800 ºC for 1 h[21,22].

Three boron (B) powders[23] of different grain sizes (0.05 μm, 0.20 μm and 2.00 μm) were used as starting materials. Commercial boron powder with 0.05 μm mean grain size was mostly amorphous boron with negligible content of β-rhombohedral phase. Boron powder with a mean grain size of 0.20 μm contains β-tetragonal boron phase due to stabilization effect of the tetragonal α-$B_{49.94}C_{1.82}$ impurity. Boron powder with mean grain size of 2.00 μm is a β-rhombohedral boron with negligible quantity of the amorphous boron phase[23].

Heating of the surface of initial boron powders has been performed in nitrogen flow in a xenon high-flux optical furnace, which is a solar simulator. The furnace involves three xenon lamps (ДКСШРБ-10000) and three ellipsoidal reflectors. Xenon tubes are centered in the focus of every ellipsoidal reflector. The calculated value of the density of the light flux energy in the focal zone is about E = 1, 4·10$^4$ kW/m$^2$ if the current in the tubes is I = 300A. Since, an emission spectrum of the xenon tubes is closely matched by that of the blackbody radiation the calculated temperature in the focal zone can be arisen up to 4000 $^0C^{21}$.

A quartz chamber was used for the processes of transformation and synthesis BN (Fig.1). Compacted samples of initial powders were tablets (diameter 20 mm and thickness 10 mm). The last were placed on a copper water-cooling screen of the quartz chamber (Fig. 2), which was positioned in the center of radiation of the three xenon emitters. Transformation and synthesis BN was carried out at the average densities of light flux energy in focal zone of set-up E ~ 0, 7 · $10^4$ kW/m$^2$. Temperature corresponds to ~ 1400 - 1900 °C. Time of the experiment was 30 min. The chamber was flowed by nitrogen. Cooper chips heated up to 500$^0$C purified the nitrogen from oxygen and other pollutions. Platelets of KOH made drying of nitrogen from the water. Additive description of origin powders and experimental were presented also in[20,24,25].

The titanium and silicon substrates were placed on a face of chamber to obtain a new structure of BN on substrate. The remaining powders collected in chamber were precipitated on the copper water-cooling screens and on a quartz surface of the chamber.

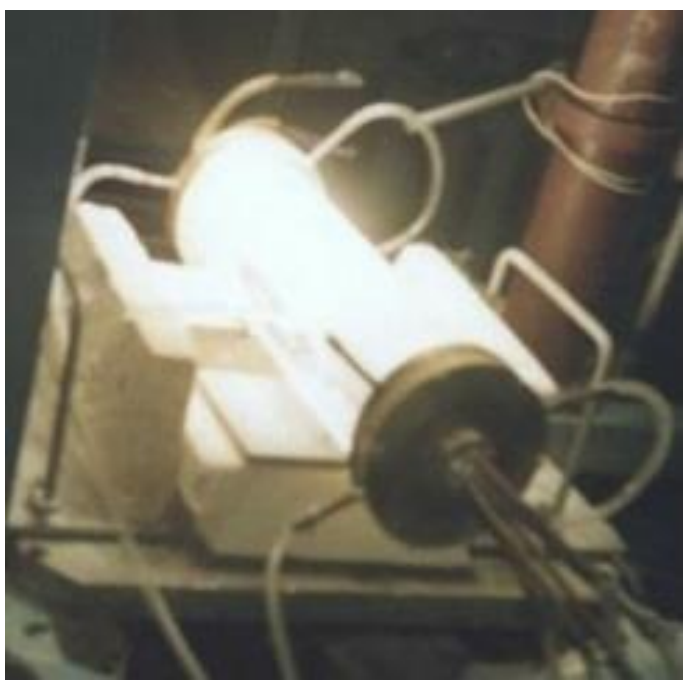 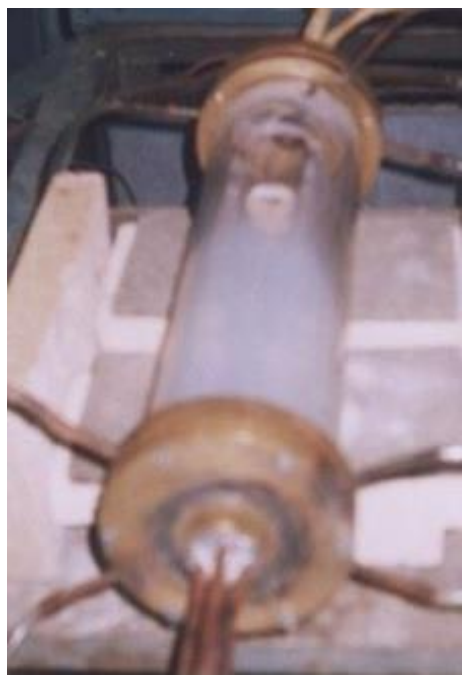

a                                                                                       b

Fig. 1. The quartz chamber for synthesis: a - in the process of synthesis; b - after the experiment.

Observation of initial and resulting powders was performed *Superprobe 733, JSM-6490* Scanning Electron Microprobes and Transmission Electron Microscope *JEM-2100F* (*JEOL* Ltd., Tokyo, *Japan*). The powders were also examined using X-ray diffraction (diffractometer "DRON-3.0", radiation of $K_\alpha$ – Cu). Raman spectroscopy has provided valuable structural information about powders. Characterization was performed with a Dilor spectrometer (XY 800) for Raman scattering in Centre of Photonics and Materials for Prospective Applications (CEPHOMA, Poland). The excitation was induced by a 2 W Argon Laser, 514.5 nm, or 488 nm and 10 nW HeNe, 632.8 nm. Dilor XY 800 Spectrograph has motorized xy stage for confocal microscope, 0.1 µm resolution, software for Raman Map.

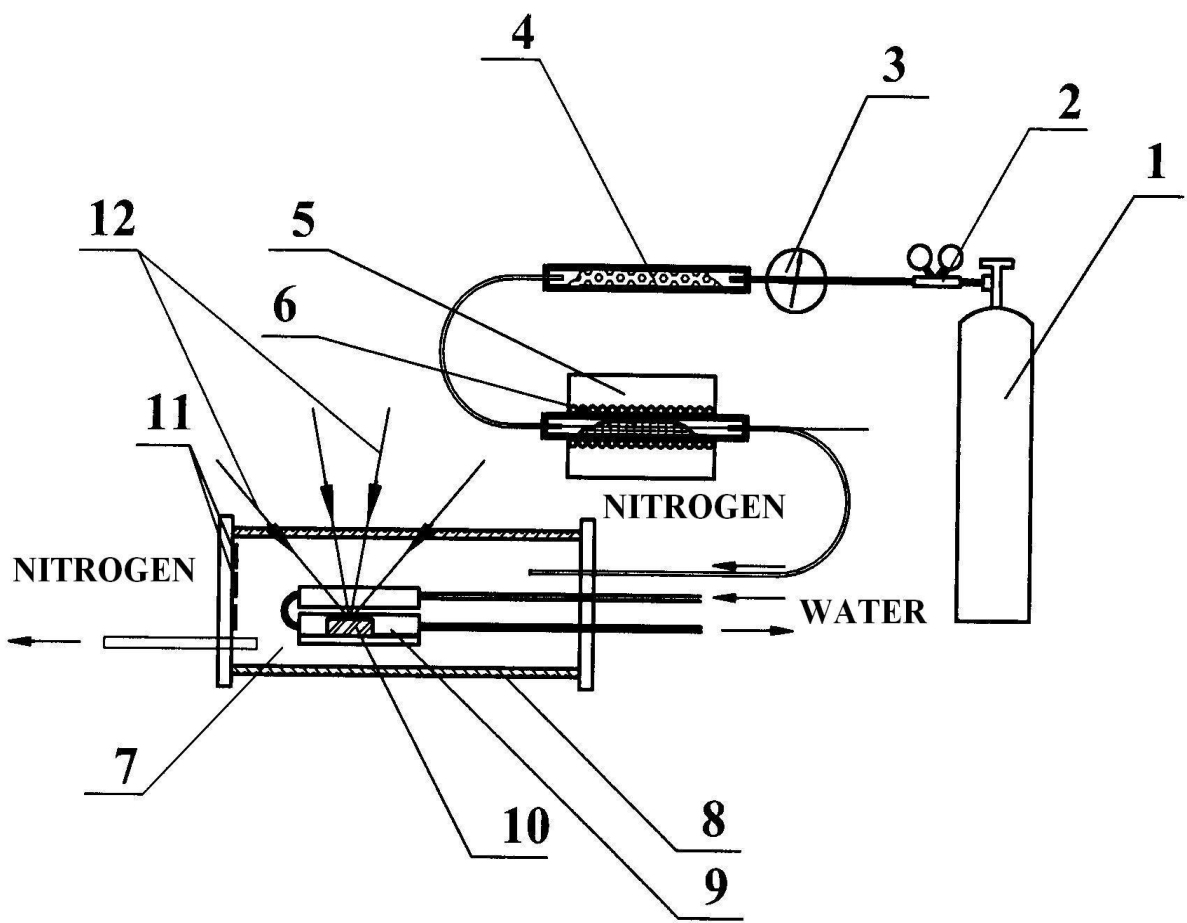

Fig. 2. The experimental sketch. Where: 1 – a pressure tank filled with gaseous nitrogen; 2 – a pressure regulator; 3 – a pressure gauge; 4 – a glass tube filled by KOH; 5 – a furnace; 6 – copper chips; 7 – a working chamber; 8 – a quartz cylinder; 9 – copper screens cooled by water; 10 – a sample of h-BN; 11– flanges cooled by water and place for substrates; 12 – concentrated light flux.

## RESULT AND DISCUSSION

Multiwalled boron nitride nanotubes (whiskers) were grown in flow of purified and dried nitrogen in the result of transformation of h-BN in the result of melting of a surface of compacted sample in condition of a xenon high-flux optical furnace at high temperatures (~1800 °C) without any catalysts (Fig. 3). Heating at higher temperatures (~2000 °C) results in formation of melted bubbles instead of nanotubes (whiskers) on the surface of a sample, which crystallized in tetragonal modification (Fig. 4 a, b). Hexagon hole for of-gassing (Fig. 5 a) and a small number of separate nanotubes (Fig. 5 b) on the surface of a bubble present a confirmation once more a "gaseous model" for nanotubes formation[21]. X-ray diffraction measurements have denoted that for all synthesized structures, besides of h-BN, a preferential formation boron-rich compounds consisting of two tetragonal phases of BN ($B_{51.2N}$ and $B_{25}N$, respectively) and tetragonal and rhombohedral phases of pure boron and amorphous phase were observed[21,24].

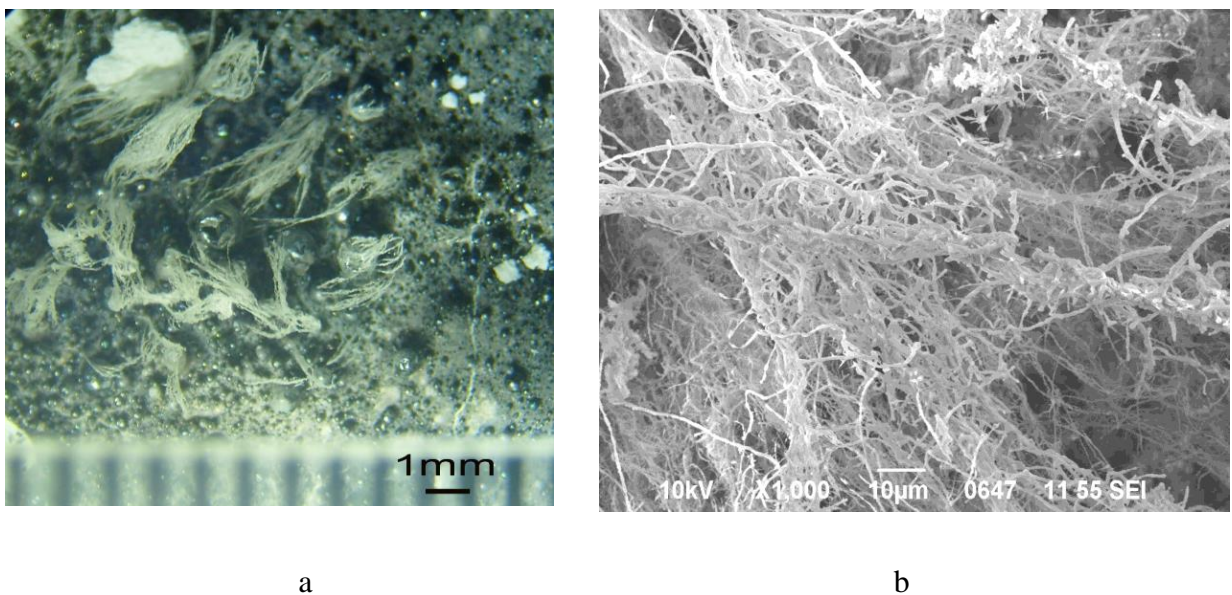

a b

Fig. 3. Optical (a) and SEM (b) image of BNNTs (whiskers) grown on the surface of compacted samples of h-BN under concentrated light heating.

It appears that TEM is not a well-adapted tool for study of boron nitride nanostructures because they are not stable to decomposition under an electron beam. Nitrogen termination[26] and point defects[27] forming under electron irradiation have been suggested. It was shown that these defects might govern the electronic and magnetic properties of h-BN systems[26]. Therefore, TEM study can display only a degraded image of graphene-like BN nanostructures (Fig. 6 a) which was forming at high temperatures without any catalysts. BN nanotube coated nonhomogeneous thick shell has a complicated structure (Fig. 6 b). The last composed of polycrystalline nanostructures according to electron diffraction pattern taken from its surface[21,22,24]. Appearance of shell can be explained by the formation of mixes of various boron nitride nanocrystalline phases. Strong chemical bonding between atoms in a given layer and weak interlayer interaction in layered boron nitrides specify an opportunity of physical and chemical intercalations by various atoms and molecules[28]. The presence of graphene-like BN nanostructures along with nanotubes does not contradict to a "gaseous model" for nanotubes formation[21], since, during a heating a bubble in the process of-gassing can draw out or blow up forming of graphene-like nanostructures in depend on temperature distribution.

Raman spectroscopy is a popular nondestructive, ambient probing tool to characterize the structure and usually imposes very little constraint on the substrate size. Due to the inhomogeneity of our synthesis samples, recording their Raman spectra was a very problematic. These spectra can be understood by comparing them with spectra obtained on high-purity commercial samples (Fig. 7).

The spectrum of h-BN is very similar to that for BN nanotubes (whiskers) which were identified by SEM and TEM (Fig. 3, 6 b). The dominant feature of the spectra is a peak at 1358 cm$^{-1}$ for h-BN and 1360 cm$^{-1}$ for BN nanotubes (whiskers) that corresponds to $E_{2g}$ mode of h-BN in agreement with previous works[29].

The peak in the BNNT's spectrum is shifted to higher frequencies by 2 cm$^{-1}$ and it is broadened asymmetrically because a half-width of the peaks for h-BN and BN nanotubes (whiskers) is 11 and 21 cm$^{-1}$ respectively. There are relationships between the h-BN frequency shift, the broadening of the mode, and the particle size. An upshift and broadening of the $E_{2g}$ mode are typical for small h-BN crystallites[30]. This conclusion confirms the results of TEM study which suggests that nanotubes coated nonhomogeneous thick shell of polycrystalline nanostructures of h-BN.

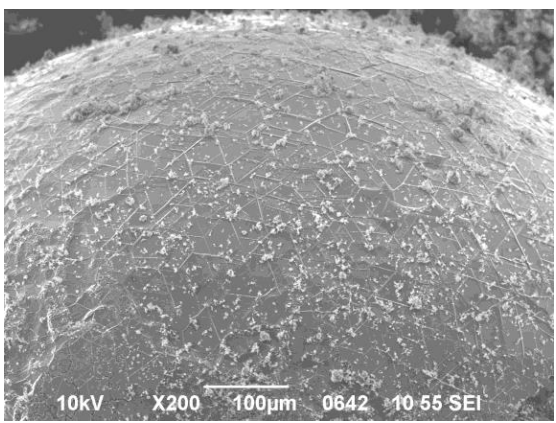 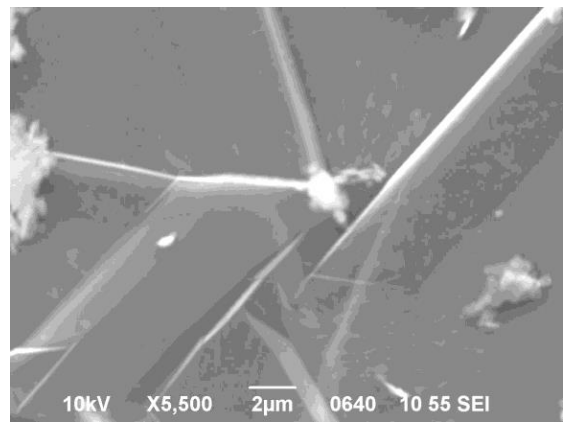

a  b

Fig. 4. SEM image of a bubble grown on the surface of compacted samples of h-BN under concentrated light heating at magnification: a – 200x; b – 5500x.

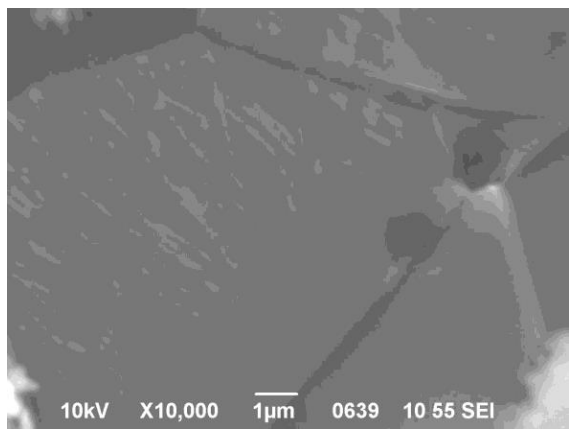 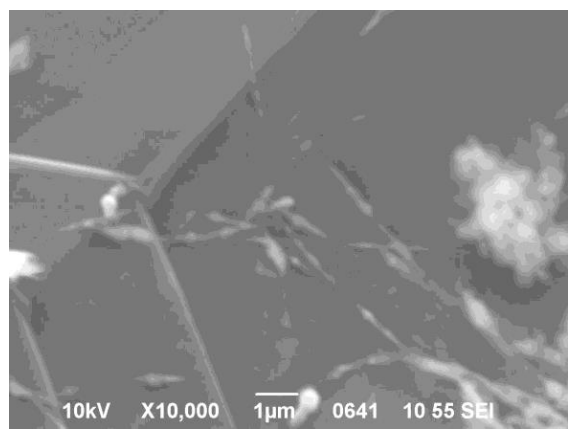

a  b

Fig. 5. SEM image of the surface of a bubble where hexagon hole for gas output (a) and a small number of separate nanotubes (b) confirm an effect of gases on nanotubes formation.

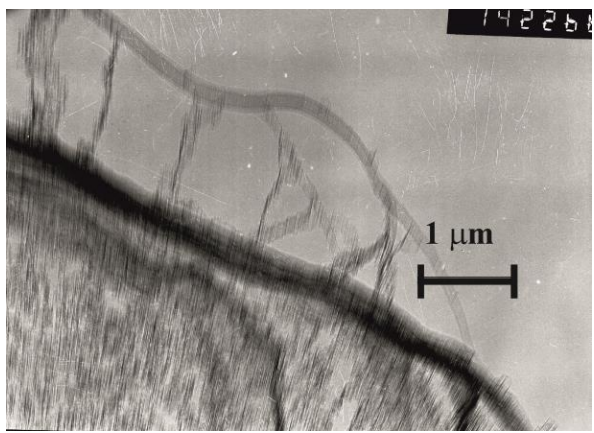 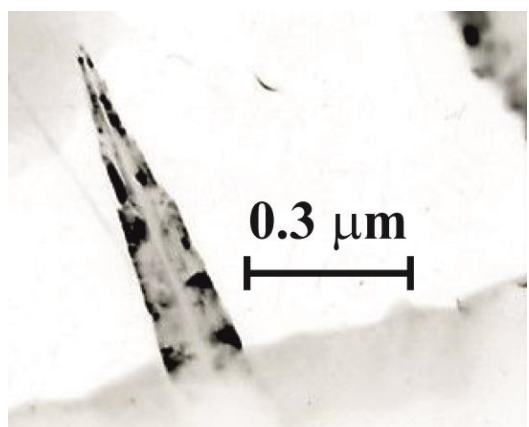

a  b

Fig. 6. TEM image of graphene-like BN nanostructures (a) and separate BN tube with a thick polycrystalline shell (b).

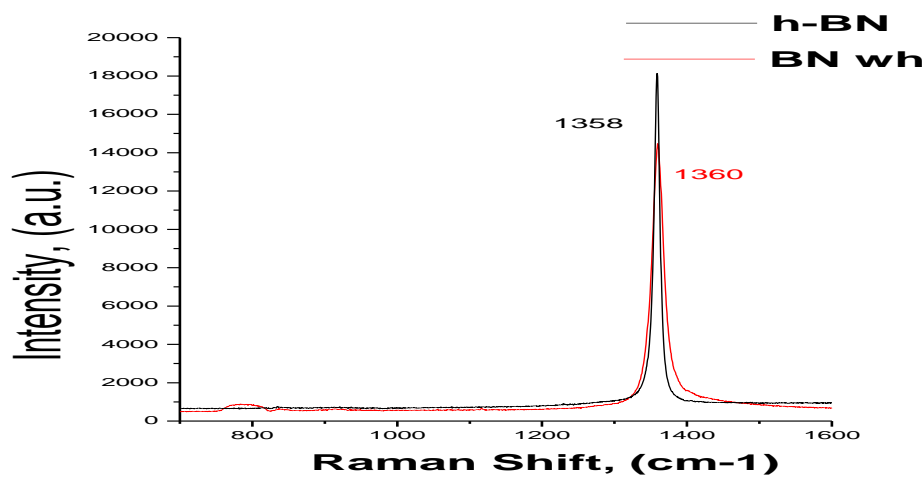

Fig. 7. Raman spectra of initial h-BN powder and synthesized BN nanotubes (whiskers).

Compared to an infinitely long system with periodic boundary conditions like h-BN, the force constant for nanotubes (whiskers) could not be reduced because of size effects. Given that the nanotubes (whiskers) have a broader size distribution, the force constants, which are size dependent, will also have a broader distribution, which in turn leads to a larger frequency range (Fig. 7). The shift and broadening can be explained in a formulation of the Raman cross section for scattering from nanocrystals in which the wave-vector uncertainty of the phonons is related to the crystal grain size. Therefore, the grain size of synthesized nanostructures is lower than grain size of initial powders.

Three absorption frequency regimes for BN nanotubes at ~809, ~1369, and ~1545 cm$^{-1}$ were considered in[31]. The weak absorption peak at ~809 cm$^{-1}$ is associated with the out-of-plane radial buckling mode where boron and nitrogen atoms are moving radially inward or outward. A detailed analysis of our Raman spectra of BN nanotubes (whiskers) (Fig. 8) displays three peaks at 788, 838 and 915 cm$^{-1}$ instead of this peak. Supposedly, splitting of the radial buckling vibration is related to the complex interaction of buckling vibration of different h-BN sheets that composed of polycrystalline shell[28].

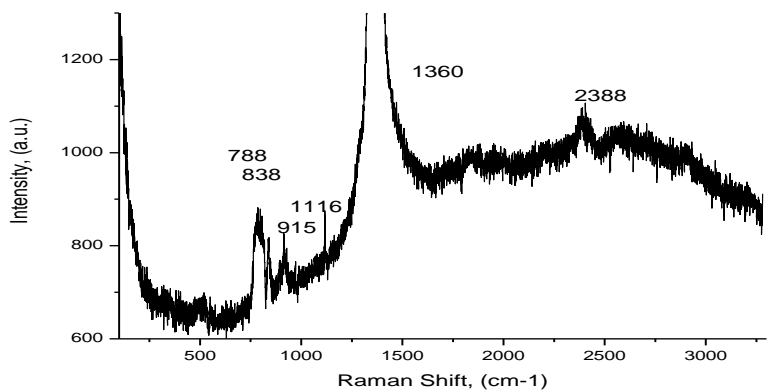

Fig. 8. A detailed analysis of a Raman spectrum of BN nanotubes (whiskers).

Presence of a liquid phase and of-gassing in the process of heating is very important for nanotubes (whiskers) formation from the initial h-BN powder at high temperatures under concentrated light[21]. It was proposed to perform a surface modification of an initial powder by a sulfuric acid. The initial h-BN powder has reacted with sulfuric acid at room temperature due to

presence of impurity of $B_2O_3$ on the surface of nanosized powder according to reaction:

$$B_2O_3 + 3\ H_2SO_4 = B_2(SO_4)_3 + 3\ H_2O \tag{1}$$

Water vapor contributes to growing of new BN plates under concentrated light heating which is not differ from plates of initial powder (Fig. 9). Supposed that of-gassing of water vapor from the modified initial powder of h-BN will stipulate easier nanotubes (whiskers) formation. However, the modification of h-BN by $H_2SO_4$ results in formation of low dense bubbles on the surface enriched with oxygen before heating (Fig. 10 a). The process of the bubbles drawing out was not observed during heating. New bubbles blow up and burst forming thick BN films (Fig. 10 b).

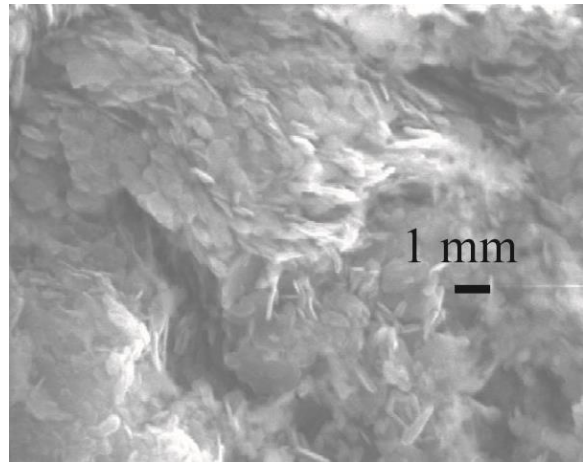

Fig. 9. SEM image of BN plates grown from modified h-BN under concentrated light heating.

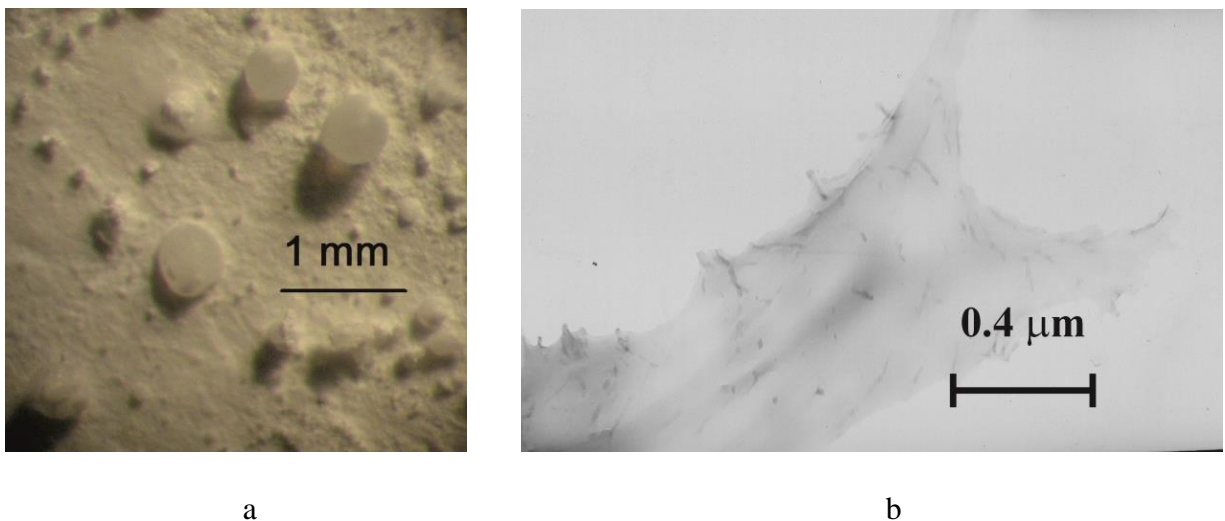

a                                                                                           b

Fig. 10. A micrograph of bubbles formed on the surface of modified by $H_2SO_4$ sample of h-BN before heating (a) and TEM image of a film grown under concentrated light heating (b).

Direct synthesis of BN using initial boron powders of mean sizes: 0,05 μm, 0,20 μm and 2,00 μm in flow of nitrogen with impurity of $H_2O$ results in formation of equiaxed, plate-like and film-like structures of different sizes[23]. The most active boron powder of mean size 0,05 μm stipulates formation large film-like structures with an area of about 4,00 μm$^2$ and mainly composed of 76 % sassolite $H_3BO_3$ (boric acid) and 33% BN (Fig. 11). Presence of carbon impurity in boron powder of mean size 0,20 μm results in formation equiaxed nanosized powders of the smallest mean size 0,30 μm[23]. Synthesis of BN using initial boron powders of mean size of 0,05 μm in flow of purified and dried nitrogen promotes formation of equiaxed and plate-like structures of different sizes (Fig. 12).

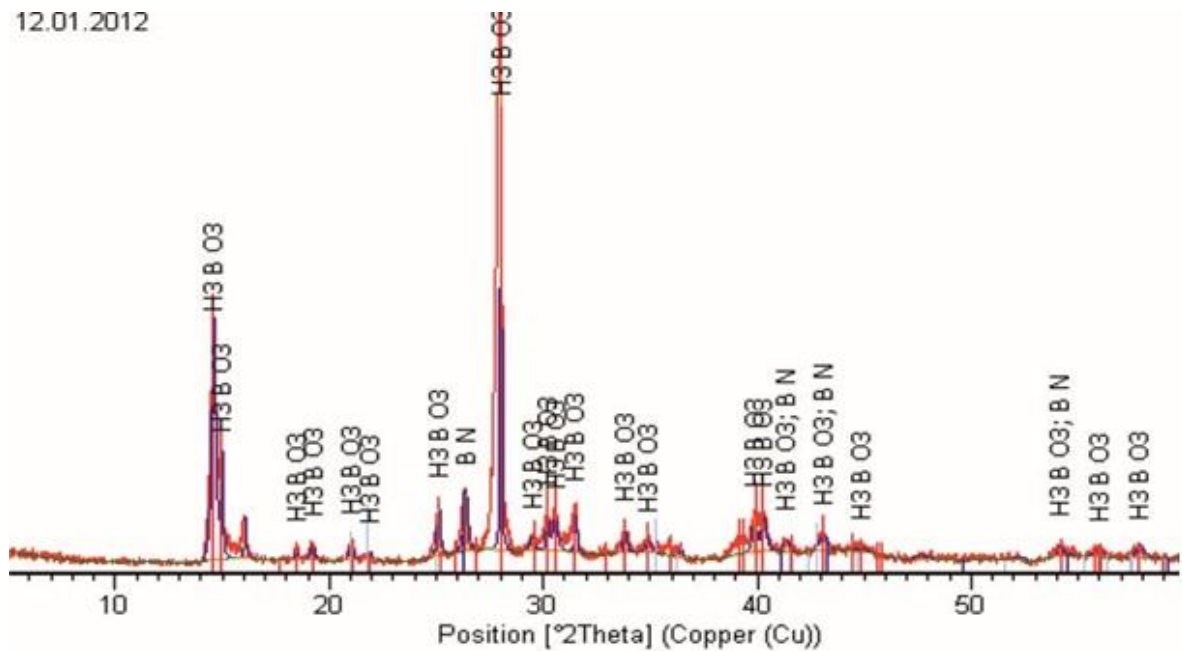

Fig. 11. Phase composition of BN powder produced from boron of mean sizes of 0,05 μm in presence of $H_2O$.

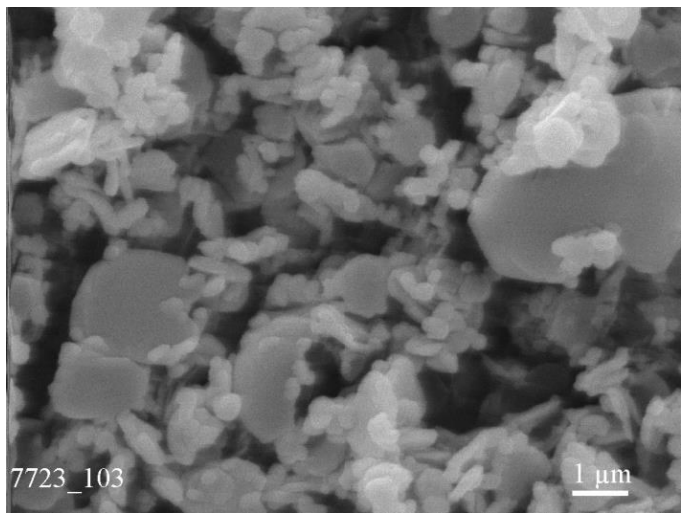

Fig. 12. SEM image of BN produced from initial boron powder of particle size of 0,05 μm in flow of purified and dried nitrogen.

Incorporation $NiSO_4$ as catalyst into initial powder of boron with a grain size of 0,2 μm promotes formation of BN nanostructures with a particle size considerably smaller than those of original powder (Fig. 13 a). The changes in morphology and in size of these nanostructures were defined by the place of their deposition relative to the reaction zone (Fig. 13).

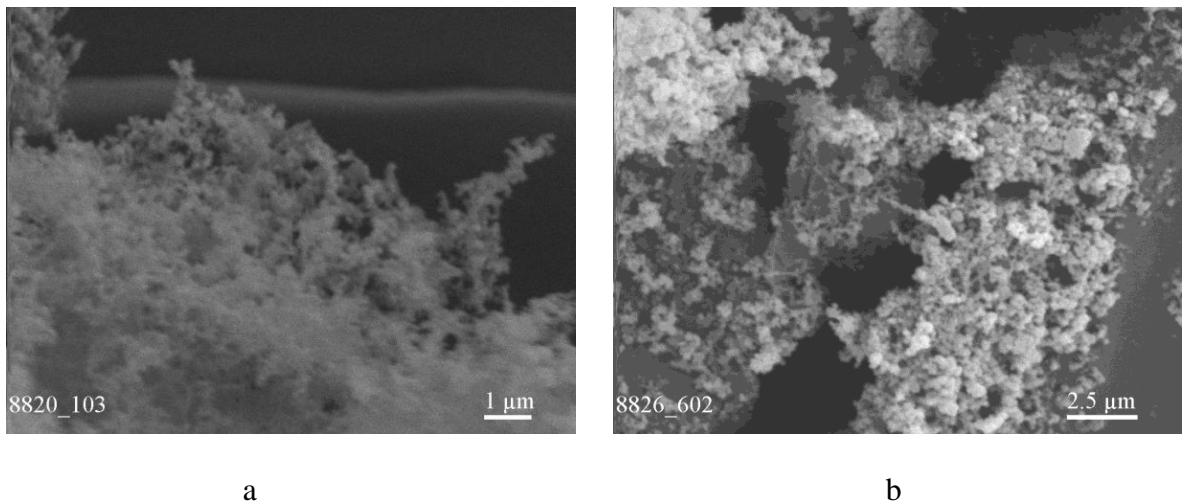

a  b

Fig. 13. SEM image of BN produced from initial boron powder of particle size of 0,2 μm in the presence of NiSO₄. The powder material was deposited: a – near a reaction zone; b – at the longer distance from the reaction zone.

## CONCLUSION

Heating in a xenon high-flux optical furnace has a number of positive features and benefits, which permit to produce structures of different morphology. High-temperature gradients promote formation of BN nanotubes (whiskers) around a reaction zone in flow of dried and purified nitrogen. Increase of temperature stipulates their growth on the surface of melted bubbles. Nanotubes (whiskers) have complicated structure and preferentially enriched by boron. Presence of a water vapor contributes to graphene-like structures making. Increase of a water vapor facilitates an increase of a thickness of a film and formation of sassolite ($H_3BO_3$). Incorporation $NiSO_4$ as catalyst into initial powder of boron promotes formation of smaller BN nanostructures. "Gaseous model" developed for nanotubes logically explains a graphene-like structure formation.

## Acknowledgements

We acknowledge support of CRDF (UKE2-7034-KV-11), TÜBİTAK and JSPS. We acknowledge also support from Prof. Hirofumi Takikawa, Prof. Mototsugu Sakai, and Associate Prof. Hiroyuki Muto.

## REFERENCE

1. Ansaloni, L. M. S. & de Sousa, E. M. B. Boron Nitride Nanostructured: Synthesis, Characterization and Potential Use in Cosmetics. *Mater. Sci. Appl.* **04,** 22–28 (2013).

2. Moussa, G. *et al.* Nanostructured Boron Nitride: From Molecular Design to Hydrogen Storage Application. *Inorganics* **2,** 396–409 (2014).

3. Zhi, C., Bando, Y., Tang, C. & Golberg, D. Boron nitride nanotubes. *Mater. Sci. Eng. R Reports* **70,** 92–111 (2010).

4. Kuzhir, P. P. *et al.* Boron Enriched Unfired Phosphate Ceramics as Neutron Protector. *Nanosci. Nanotechnol. Lett.* **4,** 1104–1109 (2012).

5. Nag, A. *et al.* Graphene analogues of BN: Novel synthesis and properties. *ACS Nano* **4,** 1539–1544 (2010).

6. Han, W., Bando, Y., Kurashima, K. & Sato, T. Formation of Boron Nitride (BN) Fullerene-Like Nanoparticles and (BN)xCy Nanotubes Using Carbon Nanotubes as Templates. *Jpn. J. Appl. Phys.* **38,** L755–L757 (1999).

7. Chen, Y., Conway, M., Williams, J. S. & Zou, J. Large-quantity production of high-yield boron nitride nanotubes. *J. Mater. Res.* **17,** 1896–1899 (2011).

8. Bengu, E. & Marks, L. D. Single-walled BN nanostructures. *Phys. Rev. Lett.* **86,** 2385–2387 (2001).

9. Lourie, O. R. *et al.* CVD Growth of Boron Nitride Nanotubes. 1808–1810 (2000).

10. Komatsu, S., Shimizu, Y., Moriyoshi, Y., Okada, K. & Mitomo, M. Preparation of boron nitride nanocapsules by plasma-assisted pulsed laser deposition. *J. Appl. Phys.* **91,** 6181 (2002).

11. Oku, T., Kuno, M., Kitahara, H. & Narita, I. Formation, atomic structures and properties of boron nitride and carbon nanocage fullerene materials. *Int. J. Inorg. Mater.* **3,** 597–612 (2001).

12. Dibandjo, P., Bois, L., Chassagneux, F. & Miele, P. Thermal stability of mesoporous boron nitride templated with a cationic surfactant. *J. Eur. Ceram. Soc.* **27,** 313–317 (2007).

13. Chen, L. *et al.* A room-temperature approach to boron nitride hollow spheres. *Solid State Commun.* **130,** 537–540 (2004).

14. Hwang, H. J. *et al.* Boron nitride nanofibers by the electrospinning technique. *Macromol. Res.* **18,** 551–557 (2010).

15. Bernard, S., Chassagneux, F., Berthet, M. P., Cornu, D. & Miele, P. Crystallinity, crystalline quality, and microstructural ordering in boron nitride fibers. *J. Am. Ceram. Soc.* **88,** 1607–1614 (2005).

16. Pacilĺ, D., Meyer, J. C., Girit, Ç. & Zettl, a. The two-dimensional phase of boron nitride: Few-atomic-layer sheets and suspended membranes. *Appl. Phys. Lett.* **92,** 212–214 (2008).

17. Kim, J. *et al.* High Purity and Yield of Boron Nitride Nanotubes Using Amorphous Boron and a Nozzle-Type Reactor. *Materials (Basel).* **7,** 5789–5801 (2014).

18. Singhal, S. K., Srivastava, A. K. & Mathur, R. B. Growth of Boron Nitride Nanotubes Having Large Surface Area Using Mechanothermal Process. *World J. Nano Sci. Eng.* **01,** 119–128 (2011).

19. Rao, C. N. R. & Govindaraj, A. Synthesis of inorganic nanotubes. *Adv. Mater.* **21,** 4208–4233 (2009).

20. Frolov, A. A., Sartinska, L. L., Koval'A, Y. & Danilenko, N. A. Application of the optical furnace for nanosized boron nitride production. *Nanomaterials* **2,** 4 (2008).

21. Sartinska, L. L. Catalyst-free synthesis of nanotubes and whiskers in an optical furnace and a gaseous model for their formation and growth. *Acta Mater.* **59,** 4395–4403 (2011).

22. Sartinska, L. L. *et al.* Transformation of fine-grained graphite-like boron nitride induced by concentrated light energy. *Mater. Chem. Phys.* **109,** 20–25 (2008).

23. Sartinska, L. L., Eren, T., Altay, E. & Frolov, G. A. EFFECT OF MOISTURE ON THE BORON NITRIDE FORMATION FROM ELEMENTS IN A XENON HIGH FLUX OPTICAL FURNACE. 165–168

24. Sartinska, L. L. *et al.* Catalyst-free synthesis and characterization of boron nitride whiskers and nanotubes. *Mater. Lett.* **65,** 1791–1793 (2011).

25. Sartinska, L. in *Boron Rich Solids* 303–318 (Springer, 2011).

26. Kotakoski, J., Jin, C. H., Lehtinen, O., Suenaga, K. & Krasheninnikov, a. V. Electron knock-on damage in hexagonal boron nitride monolayers. *Phys. Rev. B - Condens. Matter Mater. Phys.* **82,** 1–4 (2010).

27. Zobelli, a. *et al.* Defective structure of BN nanotubes: From single vacancies to dislocation lines. *Nano Lett.* **6,** 1955–1960 (2006).

28. Chkhartishvili, L. Molar Binding Energy of Zigzag and Armchair Single-Walled Boron Nitride Nanotubes. *Mater. Sci. Appl.* **01,** 222–245 (2010).

29. Gorbachev, R. V *et al.* Hunting for monolayer boron nitride: optical and Raman signatures. *Small* **7,** 465–8 (2011).

30. Arenal, R. *et al.* Raman spectroscopy of single-wall boron nitride nanotubes. *Nano Lett.* **6,** 1812–1816 (2006).

31. Wu, J. *et al.* Raman spectroscopy and time-resolved photoluminescence of BN and BxCyNz nanotubes. *Lawrence Berkeley Natl. Lab.* (2004). at <http://escholarship.org/uc/item/96c9d6d0#page-5>